\documentclass[12pt]{iopart}

\renewcommand{\imath} {\rmi}

\usepackage{epsf,iopams,graphicx}

\begin{document}

\title[The CM equation system and the ensemble average in the Gaussian ensembles]
{The Calogero-Moser equation system and the ensemble average in the Gaussian ensembles}

\author{H.-J. St\"ockmann}

\address{Fachbereich Physik der Philipps-Universit\"at Marburg,
D-35032 Marburg, Germany}

\begin{abstract}
From random matrix theory it is known that for special values of the coupling constant
the Calogero-Moser (CM) equation system is nothing but the radial part of a generalized
harmonic oscillator Schr\"odinger equation. This allows an immediate construction of the
solutions by means of a Rodriguez relation. The results are easily generalized to
arbitrary values of the coupling constant. By this the CM equations become nearly
trivial.

As an application an expansion for $\langle \rme^{\imath{\rm Tr}(XY)}\rangle$ in terms of
eigenfunctions of the CM equation system is obtained, where $X$ and $Y$ are matrices
taken from one of the Gaussian ensembles, and the brackets denote an average over the
angular variables.
\end{abstract}


\submitto{\JPA}

\ead{stoeckmann@physik.uni-marburg.de}

\section{Introduction}

There is only a very limited number of many-body systems allowing explicit solutions
\cite{hop92}. This is why the Calogero-Moser (CM) equation system from the very beginning
has attracted much interest. In its original version the system Hamiltonian reads

\begin{equation}\label{1}
  \hat{H}=-\frac{1}{2}\sum\limits_{n=1}^N\frac{\partial^2}{\partial \hat{x}_n^2}
  + \sum\limits_{n>m}
  \left[\frac{1}{2}\left(\hat{x}_n-\hat{x}_m\right)^2+\frac{g}
  {\left(\hat{x}_n-\hat{x}_m\right)^2}\right]
  \,.
\end{equation}

The parameter $g$ describes the strength of the pair interaction potential
decreasing with the square of the distance. The eigenfunctions of $\hat{H}$,
given already by Calogero in his original work \cite{cal71}, may be written
as

\begin{equation}\label{3}
  \psi_{nk}(\hat{x})=z^{\beta/2}\varphi_{nk}(r)P_k(\hat{x})\,,
\end{equation}
where $\beta=1+\sqrt{1+4g}$, and

\begin{equation}\label{2a}
    z=\prod\limits_{n>m}\left|\hat{x}_n-\hat{x}_m\right|\,,\qquad
  r^2=\sum\limits_{n>m}
  \left(\hat{x}_n-\hat{x}_m\right)^2\,.
\end{equation}

$\hat{x}$ is a short-hand notation for $(\hat{x}_1,\hat{x}_2,\dots,\hat{x}_N)$. The
radial eigenfunctions $\varphi_{nk}(r)$ read

\begin{equation}\label{4}
  \varphi_{nk}(r)=L_n^{(b)}\left(\frac{1}{2}r^2\right)e^{-\frac{1}{2}r^2}\,,
\end{equation}
where $L_n^{(b)}(x)$ is a generalized Laguerre polynomial $(n=0,1,\dots)$,
and

\begin{equation}\label{4a}
  b=k+\frac{1}{2}(N-2)+\frac{\beta}{4}N(N-1)
\end{equation}
(there is a misprint in Eq. (2.13) of Ref. \cite{cal71} for $b$). $P_k(\hat{x})$ is a
polynomial of degree $k$, totally symmetric in the variables $\hat{x}_n$. It is obtained
as a solution of

\begin{equation}\label{5}
  \left[\sum\limits_n\frac{\partial^2}{\partial \hat{x}_n^2}+ \beta
  \sum\limits_{n\ne m}\frac{1}{\hat{x}_n-\hat{x}_m}\frac{\partial}{\partial \hat{x}_n}
  \right]P_k(\hat{x})=0\,.
\end{equation}

The eigenvalues corresponding to $\psi_{nk}(\hat{x})$ are given by

\begin{equation}\label{5a}
  E_{nk}=\frac{1}{2}(N-1)+\frac{\beta}{4}N(N-1)+2n+k\,.
\end{equation}

Apart from a shift of the energy zero and differing multiplicities of the energy levels
this is just the spectrum of an harmonic oscillator.

An equivalent form for the Hamiltonian, which is more convenient in the
present context, is given by

\begin{equation}\label{1a}
  H=\frac{1}{2}\sum\limits_{n=1}^N\left(-\frac{\partial^2}{\partial x_n^2}+x_n^2\right)
  -\frac{\beta}{2}\sum\limits_{n\ne m}\frac{1}{x_n-x_m}\frac{\partial}{\partial x_n}\,.
\end{equation}
$\hat{H}$ is obtained from $H$ by introducing centre of mass variables
$\hat{x}_n =x_n-\frac{1}{N}\sum_n x_n$, and applying the transformation

\begin{equation}\label{2}
  \hat{H}=z^{-\beta/2}H\,z^{\beta/2}-H_{\rm c.m.}\,,
\end{equation}
where $H_{\rm c.m.}$ is the centre of mass Hamiltonian.

From the very beginning there was the strong suspicion that there must be a close
connection between the CM system and the quantum mechanics of the harmonic oscillator,
just from the inspection of the spectrum (\ref{5a}). But it lasted for 30 years, until
the complete equivalence of the two systems was proven by Gurappa, Panigrahi
\cite{gur99}. For the proof a quite evolved operator technique was applied. There were
many futile attempts as well to find expressions for the polynomial $P_k(\hat{x})$, until
recently an explicit construction of the solutions of the CM system has been achieved by
Ujino, Wadati \cite{uji94}.

In this paper it will be demonstrated that there is a much simpler connection between the
quantum mechanics of the harmonic oscillator and the CM system. In fact for the special
cases $\beta=1,2,4$ this connection will show up to be completely trivial.

The key ingredient to this result comes from random matrix theory, where the CM equations
enter via the calculation of averages in the Gaussian ensembles (see e.\,g. Refs.
\cite{guh02a, nis03}). Since this background probably is not known to all of the experts
engaged in the CM equations, a short recapitulation of the aspects relevant in the
present context is given in Section \ref{RM1}. In the subsequent Section \ref{CM1} the
equivalence of the CM system with the Schr\"odinger equation of the harmonic oscillator
is established for the cases $\beta=1,2,4$. This result is generalized to arbitrary
values of $\beta$ in Section \ref{CM2}. In Section \ref{RM2} an application to the
calculation of averages in the Gaussian ensembles is presented. The paper ends with a
short discussion in Section \ref{DI}.

\section{The averaging problem in the Gaussian ensembles}\label{RM1}

An issue of central importance in random matrix theory is the calculation of averages
over products and ratios of spectral determinants,

\begin{equation}\label{6}
  M_\nu(E_i)=\left<\prod\limits_i\left|E_i-H\right|^{\nu_i}\right>\,,
\end{equation}

where the matrix elements of $H$ are assumed to be Gaussian distributed, and the $\nu_i$
may take positive and negative integer values. The standard way to calculate such
averages uses supersymmetry techniques. In this approach after some steps an integral of
the type

\begin{equation}\label{7}
  I(Y)=\int \rmd [x] \rme^{\imath{\rm Tr}(XY)-{\rm Tr}[F(X)]}
\end{equation}
is met, where $X$, $Y$ are  $N\times N$ supermatrices. The integral is both over the
symmetric and the antisymmetric components of $X$ (see Ref. \cite{ver85a} for more
details). Diagonalizing $X$,

\begin{equation}\label{8}
  X=RX_DR^{-1}\,,
\end{equation}
where $X_D={\rm diag}(x_1,\dots,x_N)$, one has
\begin{equation}\label{9}
  I(Y)=\int \left(\prod\limits_i \rmd x_i\,\rme^{-F(x_i)}\right) \,B(x)f(x,y)\,.
\end{equation}

$B(x)$ is the radial part of the functional determinant, and $f(x,y)$ is given by

\begin{equation}\label{10}
  f(x,y)=\left<\rme^{\imath{\rm Tr}(XY)}\right>\,,
\end{equation}
where the brackets denote the average over the angular variables. From the invariance
property of the trace is is obvious that $f(x,y)$ depends only on the eigenvalues of $X$
and $Y$, but not on the respective angular variables.

The practical applicability of supersymmetry techniques relies on convenient expressions
for $f(x,y)$.  For Hamiltonians taken from the Gaussian unitary ensemble (GUE), the
angular average has been calculated by Itzykson and Zuber \cite{itz80} for ordinary
matrices. Their result has been generalized to supermatrices by Guhr \cite{guh91}.
Explicit expressions for the average (\ref{6}) over spectral determinants have been given
by Fyodorov, Strahov \cite{fyo03b}.

For the Gaussian orthogonal ensemble (GOE), much more important from the practical point
of view, and the Gaussian symplectic ensemble (GSE) the situation is unsatisfactory.
Expressions for the ensemble average of the negative moments of the spectral determinant
have been given by Fyodorov, Keating \cite{Fyo03a}. Brezin and Hikami achieved to express
averages over products of spectral determinants in terms of complicated integrals over
determinants of quaternionic matrices \cite{bre01}. Recursive relations for $f(x,y)$ with
respect to the rank $N$ have been given by Guhr and K\"ohler \cite{guh02a,guh02b}.

For all ensembles there are expansions of the type

\begin{equation}\label{11}
  f(x,y)=\sum\limits_n a_n^{2/\beta} P_n^{2/\beta}(x)P_n^{2/\beta}(y)\,,
\end{equation}
where the $P_n^{2/\beta}(x)$ are Jack polynomials \cite{mui82,oko97}. The sum is over all
partitions $n=(n_1,\dots,n_N)$, where $|n|=\sum_in_i=n$ is the degree of the polynomial.
The polynomials are totally symmetric in the variables, and are orthogonal on the
interval $[-\frac{1}{2},\frac{1}{2}]$ with the weight function
$\prod\limits_{n>m}\left|x_n-x_m\right|^\beta$,

\begin{equation}\label{12}
  \int\limits_{-1/2}^{1/2}\rmd x_1\,\cdots\int\limits_{-1/2}^{1/2}\rmd x_n
 \prod\limits_{n>m}\left|x_n-x_m\right|^\beta
 P_n^{2/\beta}(x)P_m^{2/\beta}(x)
 \sim\delta_{nm}\,,
\end{equation}
see e.\,g. Ref. \cite{bor03}.

Unfortunately, the results give in the mentioned references are not very convenient for
the practical application. This is why for the GOE, in contrast to the GUE, only a quite
small number of analytic results on ensemble averages is available. On the other hand
there are a number of experiments, in particular in chaotic microwave cavities performed
in the author's group, where exactly such averages are needed \cite{sch01d,men03}. This
was the original motivation for this work.

\section{The Calogero-Moser equations for $\bbeta{\bf=1,2,4}$}\label{CM1}

$f(x,y)$, considered as a function of $x$, obeys the differential equation

\begin{equation}\label{13}
  \left(\Delta+{\rm Tr}Y^2\right)f(x,y)=0\,,
\end{equation}
where

\begin{equation}\label{17a}
  {\rm Tr}Y^2=\sum_{ij} Y_{ij}Y_{ji}=\sum_n y_n^2,
\end{equation}
and

\begin{equation}\label{17b}
  \Delta= \sum_{ij}\frac{\partial^2}{\partial x_{ij}\partial x_{ji}}\,,
\end{equation}
is a generalized Laplace operator. Since $f(x,y)$ is independent on the angular variables
of $X$, we may restrict the Laplace operator to its radial part. It is well-known from
numerous papers on random matrix theory (see e.\,g. the references cited above), how to
separate the Laplace operator into its radial and angular part, but for the convenience
of readers not familiar with the topic the essential steps shall be repeated.

To keep the discussion simple we assume that all operators are Hermitian, i.\,e. both
summation indices $i,j$ in equations (\ref{17a}) and (\ref{17b}) run from 1 to $N$. The
generalization to the two other symmetry classes is straightforward though some care has
to be taken to avoid double counting. For symmetric operators e.\,g. in all traces the
summations have to be restricted to $i\le j$. In addition we restrict ourselves to
ordinary commuting variables. The extension to supersymmetric variables, just as it was
demonstrated in Ref. \cite{guh91} for the GUE, is straightforward as well.

We start with the somewhat more general equation

\begin{equation}\label{17}
  \frac{1}{2}\left(-\Delta+ {\rm Tr}X^2\right)\psi_n(X)=E_n\psi_n\,,
\end{equation}
which is nothing but the Schr\"odinger equation for an independent superposition of
harmonic oscillators. Its solution may thus by be expressed in terms of products of
single harmonic oscillator eigenfunctions, or, using Rodriguez' formula for the Hermite
polynomials, as

\begin{eqnarray}\label{18}
  \psi_{n_{11}\dots n_{NN}}(X)&\sim& e^{\frac{1}{2}{\rm Tr}X^2}\left[\prod\limits_{ij}
  \left(-\nabla_{ij}\right)^{n_{ij}}\right]  e^{-{\rm Tr}X^2}\nonumber\\
   &\sim&\left[\prod\limits_{ij}\left(x_{ij}-\nabla_{ij}\right)^{n_{ij}}\right]
  e^{-\frac{1}{2}{\rm Tr}X^2}\,.
\end{eqnarray}
The corresponding corresponding eigenenergies are

\begin{equation}\label{19}
  E_n=\sum\limits_{ij}\left(n_{ij}+\frac{1}{2}\right)=\sum\limits_{ij}n_{ij}+\frac{N^2}{2}\,.
\end{equation}

The sub-set of solutions, depending on the radial variables only, is obtained from Eq.
(\ref{18}) as

\begin{equation}\label{20}
  \psi_n(x)\sim\left(\prod\limits_iA_+^{n_i}\right)
  e^{-\frac{1}{2}{\rm Tr}X^2}\,,
\end{equation}
where $n$ is an abbreviations for $(n_1,n_2,\dots)$, and

\begin{equation}\label{20a}
  A_+^k={\rm Tr}\left[\left(X-\nabla\right)^k\right]_{\rm rad}
\end{equation}
is a generalized creation operator. The eigenenergies corresponding to $\psi_n(x)$ are
given by

\begin{equation}\label{21}
  E_n=\sum\limits_in_i+\frac{N^2}{2}= |n|+\frac{N^2}{2}\,
\end{equation}
where $|n|=\sum_in_i$. Each partition $(n_1,n_2,\dots)$ with the same $|n|$ thus yields
an independent eigenfunction to the same eigenvalue.

Only by using invariance properties of the trace we achieved to construct the radially
symmetric solutions (\ref{20}) of the generalized harmonic oscillator Schr\"odinger
equation (\ref{17}). These functions at the same time must be solutions of the radial
part of the Schr\"odinger equation. Thus the radial part of the generalized Laplace
operator entering equation (\ref{17}) has to be determined. For the eigenfunctions in
addition we need expressions for the radial part of ${\rm Tr}(X-\nabla)^k$.

Details of the calculation can be found in \ref{app1}. For the radial representation of
the generalized creation operator we obtain

\begin{equation}\label{21a}
  A_+^k=  \sum\limits_{nm}\left[(X_D-D)^k\right]_{nm}\,,
\end{equation}
where

\begin{equation}\label{30}
  D_{nm}=\delta_{nm}\left(\frac{\partial}{\partial x_n}+\frac{\beta}{2}{\sum\limits_k}'
  \frac{1}{x_n-x_k}\right)-\frac{\beta}{2}\frac{1-\delta_{nm}}{x_n-x_m}\,.
\end{equation}

Equation (\ref{30}) holds for all Gaussian ensembles with $\beta=1,2,4$ for the GOE, the
GUE, and the GSE, respectively. For the radial part of the Laplace operator we get

\begin{equation}\label{31}
  \Delta_{\rm rad}=\sum\limits_n\frac{\partial^2}{\partial x_n^2} +
  \beta\sum\limits_{n\ne m}\frac{1}{x_n-x_m}\frac{\partial}{\partial x_n}\,.
\end{equation}

Note that this is exactly the operator entering equation (\ref{5}) for the polynomials
$P_k(x)$. For the radial part of the generalized oscillator Schr\"odinger equation
(\ref{17}) we have

\begin{equation}\label{33}
 H\psi_n(x)=E_n\psi_n(x)\,,
\end{equation}
where

\begin{equation}\label{34}
  H= \frac{1}{2}\sum\limits_{n=1}^N\left(-\frac{\partial^2}{\partial x_n^2}
 +x_n^2\right)-\frac{\beta}{2}\sum\limits_{n\ne m}\frac{1}{x_n-x_m}\frac{\partial}{\partial
 x_n}\,.
\end{equation}

This is identical with the CM Hamiltonian (\ref{1a}).

The derivation of this section has shown that for the special cases $\beta=1,2,4$ the CM
equation system is completely trivial. It is nothing but the radial part of the
generalized harmonic oscillator Schr\"odinger equation (\ref{17}). As a consequence the
solutions are also obtained in a trivial way just by picking out all linear combinations
of products of Hermitian polynomials which are totally symmetric in the variables.

\section{The general case}\label{CM2}

For $\beta=1,2,4$ the problem is thus completely solved. But all results can immediately
be transferred to arbitrary values of $\beta$.

First we notice that

\begin{equation}\label{36}
  \psi_0(x)\sim\exp\left(-\frac{1}{2}\sum_n x_n^2\right)
\end{equation}
is the ground state eigenfunction of the Calogero-Moser equation (\ref{33}) for all
values of $\beta$, with a ground state energy given by

\begin{equation}\label{37}
  E_0= \frac{N}{2}+\beta\frac{N(N-1)}{4}\,.
\end{equation}

Second, the generalized creation operators obey the commutation rule

\begin{equation}\label{32a}
  HA_+^k=A_+^k\left(H+k\right)\,,
\end{equation}
which is the generalization of a well-known relation for the one-dimensional harmonic
oscillator (see \ref{app2}). For $\beta=1,2,4$ this follows trivially from equation
(\ref{20a}). But the rule holds for arbitrary values of $\beta$, if only $A_+^k$ is
calculated from expression (\ref{21a}), which is well-defined for all values of $\beta$,
and not from equation (\ref{20a}), which looses its meaning in the general case.

Since nothing else but this commutation rule is needed for the generalized Rodriguez
relations to hold, equation (\ref{20}) can still be used to generate the eigenfunctions.
The corresponding spectrum is thus not altered by varying $\beta$, apart from a shift of
the ground state energy, a fact noticed already by Calogero \cite{cal71}.

\section{A series expansion for ${\bf\left<\rme^{\imath{\rm Tr}(XY)}\right>}$}\label{RM2}

Generalizing an expansion of $\rme^{\imath xy}$ in terms of ordinary oscillator
eigenfunctions we are now going to expand $f(x,y)$ in terms of the totally symmetric
solutions (\ref{20}) of the generalized Schr\"odinger equation (\ref{17}),

\begin{equation}\label{40}
  f(x,y)=\left<\rme^{\imath{\rm Tr}(XY)}\right>=\sum\limits_{nm}
  c_{nm}\psi_n(x)\psi_m(y)\,.
\end{equation}

In a sequence of self-explaining steps, including one integration by parts, we have

\begin{eqnarray}\label{41}
\fl \int \rmd [Y] \psi_n(y)f(x,y)
  &=&\int \rmd [Y] \psi_n(y)\left<\rme^{\imath{\rm Tr}(XY)}\right>\nonumber\\
  &=&c_n\int \rmd [Y]\left[\prod\limits_i{\rm Tr}\left(Y-\nabla_Y\right)^{n_i}
  e^{-\frac{1}{2}{\rm Tr}Y^2}\right]\rme^{\imath{\rm Tr}(XY)}\nonumber\\
  &=&c_n\int \rmd [Y] \rme^{-\frac{1}{2}{\rm Tr}Y^2}
  \left[\prod\limits_i{\rm Tr}\left(Y+\nabla_Y\right)^{n_i}e^{\imath{\rm
  Tr}(XY)}\right]\nonumber\\
  &=&c_n\,\imath^{|n|}\left[\prod\limits_i{\rm Tr}
  \left(X-\nabla_X\right)^{n_i}\right]
  \int \rmd [Y] \rme^{-\frac{1}{2}{\rm Tr}Y^2}
  \rme^{\imath{\rm Tr}(XY)}\nonumber\\
  &=&c_n\,\imath^{|n|}\left[\prod\limits_i{\rm Tr}
  \left(X-\nabla_X\right)^{n_i}\right]
  \rme^{-\frac{1}{2}{\rm Tr}X^2}\nonumber\\
&=& (2\pi)^{\frac{N^2}{2}}\imath^{|n|} \psi_n(x)\,,
\end{eqnarray}

where $c_n$ is the normalization facto of the $\psi_n(x)$. The differential is given,
either in Cartesian or radial-angular coordinates, by

\begin{equation}\label{41a}
  \rmd [Y]=\prod\limits_{ij}\rmd Y_{ij}=\prod\limits_{i< j}\left|y_i-y_j\right|^\beta
  \prod\limits_i\rmd y_i\,\rmd\Omega_\beta\,,
\end{equation}
where $\rmd\Omega_\beta$ is the angular part of the differential. Comparison with
equation (\ref{40}) yields

\begin{equation}\label{42}
  c_{nm}= (2\pi)^{\frac{N^2}{2}}\imath^{|n|} \left(\rho^{-1}\right)_{nm}\,,
\end{equation}
where $\rho$ is the matrix with the elements

\begin{equation}\label{43}
  \rho_{nm}=\int \rmd [X] \psi_n(x)\psi_m(x)\,.
\end{equation}
Eigenfunctions belonging to different $|n|$ are orthogonal, but for degenerate
eigenfunctions this is not necessarily the case. The matrix of expansion coefficients
$c_{nm}$ is thus block-diagonal, where each block corresponds to a given value of $|n|$.

The problem of diagonalization of the eigenfunctions belonging to the same $|n|$  has
been solved by Ujino, Wadati in a series of papers \cite{uji96a,uji96b,uji97}. The
authors introduced a new type of totally symmetric polynomials $j_n^{2/\beta}(x)$ they
called hidden Jack polynomials (or in short, Hi-Jack polynomials), which are obtained
from linear combinations of the $\psi_n(x)$ introduced in equation (\ref{20}). The
Hi-Jack polynomials obey the orthogonality relation

\begin{eqnarray}\label{44}
 \int\limits_{-\infty}^\infty \rmd x_1\,\cdots\int\limits_{-\infty}^\infty \rmd x_N
 \prod\limits_{k>l}\left|x_k-x_l\right|^\beta\exp\left(-\sum_kx_k^2\right)
 j_n^{2/\beta}(x)j_m^{2/\beta}(x)
 \sim\delta_{nm}\,.\nonumber\\
\end{eqnarray}

Collecting the results we obtain an expansion of $f(x,y)$ in terms of Hi-Jack
polynomials,

\begin{equation}\label{45}
  f(x,y)=\left<e^{\imath{\rm Tr}(XY)}\right>=(2\pi)^{\frac{N^2}{2}}\sum\limits_n\imath^{|n|}
  \psi_n^{2/\beta}(x)\psi_m^{2/\beta}(y)\,,
\end{equation}
where the sum is over all partitions $n$, and

\begin{equation}\label{46}
  \psi_n^{2/\beta}(x)=c_n^{2/\beta} j_n^{2/\beta}(x)\exp\left(-\frac{1}{2}\sum_kx_k^2\right)\,.
\end{equation}

$c_n^{2/\beta}$ is a constant normalizing the integral (\ref{44}) to one for $n=m$.

Equation (\ref{45}) is an alternative to expansion (\ref{11}) in terms of ordinary Jack
polynomials. It has still to be checked whether this will a real progress in the
calculation of ensemble averages in the GOE. The formulas given in reference
\cite{uji96a} for the Hi-Jack polynomials are quite complicated, but is does not seem
hopeless that a more direct generation of the polynomials is possible, if the techniques
applied in the present work are used.

\section{Discussion}\label{DI}

Using the fact that for special values of the coupling constant the CM equation system is
nothing but the radial part of a generalized harmonic oscillator Schr\"odinger equation,
the solution could be immediately constructed in terms of totally symmetric linear
combinations of products of Hermitian polynomials. The generalization to arbitrary values
of the coupling constant then was straightforward.

None of these results is really new. The equivalence of the CM equation system with the
Schr\"odinger equation of the harmonic oscillator was suspected from the very beginning
\cite{cal71} and was proven recently \cite{gur99}. Explicit solutions of the equation
system, too, have been known for some time \cite{uji94}. In fact the solutions (\ref{20})
derived in the present work are completely equivalent to the ones given in Ref.
\cite{uji94}. The fact that the radial part of the generalized harmonic oscillator
Schr\"odinger equation reduces to a special case of the CM equations, too, has been known
from random matrix for several years.

But the experts working on the CM equations and on random matrix theory did not know much
of each other as it seems. This is probably why it remained unnoticed for such a long
time that by a combination of ingredients from both sides the CM equations become nearly
trivial.

The results have been applied to derive a new expansion of $\left<e^{\imath{\rm
Tr}(XY)}\right>$ in terms of Hi-Jack polynomials, see equation (\ref{45}). Whether this
expansion is better adopted to random matrix problems as the existing expansion
(\ref{11}) in terms of Jack polynomials has still to be explored.

\section*{Acknowledgments}

Thomas Guhr, Lund, is thanked for many valuable suggestions, and for drawing my attention
to a number of relevant references. The microwave experiments, performed in the authors
group and being the motivation for this work, were supported by the Deutsche
Forschungsgemeinschaft via several grants.

\appendix

\section{The radial part of the momentum operator}\label{app1}

For Hermitian matrices Eq. (\ref{8}) reads, written in components,

\begin{equation}\label{23}
  x_{ij}=\sum\limits_n r_{in}r^*_{jn}x_n\,.
\end{equation}

Introducing the notation

\begin{equation}\label{24}
  y_k=\left\{
  \begin{array}{ll}
    x_k\,,\qquad & k\le N \,,\\
    \alpha_k\,, & k>N\,,
  \end{array}
\right.
\end{equation}
where the $\alpha_k$ are the $N(N-1)$ angular variables, we have

\begin{equation}\label{25}
  \frac{\partial x_{ij}}{\partial y_k}=J_{k,ij}=
  \sum\limits_{nm}r_{in}r^*_{jm}\hat{J}_{k,nm}\,,
\end{equation}
where

\begin{equation}\label{26}
  \hat{J}_{k,nm}=\left\{
  \begin{array}{ll}
    \delta_{kn}\delta_{km}\,, & n=m\,, \\
    S_{k,nm}\left(x_m-x_n\right)\,,\qquad & n\ne m\,,
  \end{array}
\right.
\end{equation}
and

\begin{equation}\label{27}
  S_{k,nm}=-S_{k,mn}=\sum\limits_ir^*_{in}\frac{\partial r_{im}}{\partial \alpha_k}\,.
\end{equation}

It follows

\begin{eqnarray}\label{28}
  \nabla_{ij}=\frac{\partial}{\partial x_{ji}}
  &=&\sum\limits_k \left(J^{-1}\right)_{ij,k}\frac{\partial}{\partial
  y_k}\nonumber\\
  &=&\sum\limits_{knm} r_{in}r^*_{jm}\left(\hat{J}^{-1}\right)_{nm,k}\frac{\partial}{\partial
  y_k}\,.
\end{eqnarray}

This may be written as

\begin{equation}\label{28a}
  \nabla_{ij}=\sum\limits_{nm}r_{in}\tilde{\nabla}_{nm}r_{jm}^*\,,
\end{equation}
where

\begin{equation}\label{28b}
  \tilde{\nabla}_{nm}=\delta_{nm}\frac{\partial}{\partial x_n}
  +\left(1-\delta_{nm}\right)\left[\sum\limits_k
  \left(\hat{J}^{-1}\right)_{nm,k}\frac{\partial}{\partial
  \alpha_k}-\frac{1}{x_n-x_m}\right]\,.
\end{equation}

Using representation (\ref{28a}) for the matrix elements of $\nabla$,  the radial part of
${\rm Tr}(X-\nabla)^k$ is now easily calculated. We obtain after a number of elementary
steps,
\begin{equation}\label{29}
  A_+^k=
  \sum\limits_{nm}\left[(X_D-D)^k\right]_{nm}\,,
\end{equation}
where

\begin{equation}\label{29a}
  D_{nm}=\delta_{nm}\left(\frac{\partial}{\partial x_n}+{\sum\limits_k}'
  \frac{1}{x_n-x_k}\right)-\frac{1-\delta_{nm}}{x_n-x_m}\,.
\end{equation}

Equation (\ref{29a}) holds for Hermitian matrices, i.\,e. matrices taken from the unitary
ensemble. Generalizing the calculation to the other ensembles we obtain equation
(\ref{30}).

\section{Proof of relation (\ref{32a})}\label{app2}

The $D_{nm}$  and $\Delta_{\rm rad}$ (see equations (\ref{30}) and (\ref{31})) obey the
commutation rule

\begin{eqnarray}\label{a1}
  \left[\Delta_{\rm rad},D_{nm}\right]&=&
  \beta\,{\sum\limits_k}'\Bigg[\left(D_{nk}-D_{nm}\right)\frac{1}
  {\left(x_k-x_m\right)^2}\nonumber\\
  &&-\frac{1}{\left(x_n-x_k\right)^2}\left(D_{km}-D_{nm}\right)\Bigg]\,,
\end{eqnarray}
as can be verified directly from the definitions. The relation is true both for $n=m$ and
$n\ne m$. Introducing the matrix $S$ with elements

\begin{equation}\label{a2}
  S_{nm}=\frac{1-\delta_{nm}}{(x_n-x_m)^2}-\delta_{nm}
  {\sum\limits_k}'\frac{1}{(x_n-x_k)^2}\,,
\end{equation}
equation (\ref{a1}) may be written more concisely as

\begin{equation}\label{a3}
  \left[\Delta_{\rm rad},D_{nm}\right]=\beta[D,S]_{nm}\,.
\end{equation}

Further commutation rules following directly from the definitions are

\begin{eqnarray}\label{a4}
  \left[\Delta_{\rm rad},x_n\right]&=&-2D_{nn}\,,\\
  \left[\sum_k x_k^2,D_{nm}\right]&=& -2\delta_{nm}x_n\label{a5}\,.
\end{eqnarray}

Introducing the operator $A_+$ with the components

\begin{equation}\label{a6}
  \left(A_+\right)_{nm}=x_n\delta_{nm}-D_{nm}\,,
\end{equation}
one obtains by combining equations (\ref{a3}) to (\ref{a5})

\begin{equation}\label{a7}
  \left[H,\left(A_+\right)_{nm}\right]=\left(A_+\right)_{nm}
  -\frac{\beta}{2}\left[A_+,S\right]_{nm}\,.
\end{equation}

Now the wanted commutator (\ref{32a}) can be calculated,

\begin{eqnarray}\label{a8}
  \left[H,A_+^k\right]&=&\sum\limits_{nm}\left[H,\left[(A_+\right)^k]_{nm}\right]\nonumber\\
  &=&\sum\limits_{\nu=0}^{k-1}\sum\limits_{ijnm}[(A_+)^\nu]_{ni}
  \left[H,\left(A_+\right)_{ij}\right]
  [(A_+)^{(k-\nu-1)}]_{jm}\nonumber\\
  &=&kA_+^k-\frac{\beta}{2}\sum\limits_{\nu=0}^{k-1}
  \sum\limits_{ijnm}[(A_+)^\nu]_{ni}
  \left[A_+,S\right]_{ij}
  [(A_+)^{(k-\nu-1)}]_{j m}\,.
\end{eqnarray}

The second term on the right hand side does not contribute as is easily seen, and we are
left with equation (\ref{32a}). q.\,e.\,d.


\section*{References}

\begin{thebibliography}{10}

\bibitem{hop92}
J. Hoppe, {\em Lectures on Integrable Systems} (Springer, Berlin, 1992).

\bibitem{cal71}
F. Calogero, J. Math. Phys. {\bf 12},  419  (1971).

\bibitem{gur99}
N. Gurappa and P. Panigrahi, Phys. Rev. B {\bf 59},  R2490  (1999).

\bibitem{uji94}
H. Ujino and M. Wadati, J. Phys. Soc. Jpn. {\bf 63},  3585  (1994).

\bibitem{guh02a}
T. Guhr and H. Kohler, J. Math. Phys. {\bf 43},  2741  (2002).

\bibitem{nis03}
S. Nishigaki, D. Gangardt, and A. Kamenev, J. Phys. A {\bf 36},  3137  (2003).

\bibitem{ver85a}
J. Verbaarschot, H. Weidenm\"uller, and M. Zirnbauer, Phys. Rep. {\bf 129},
  367  (1985).

\bibitem{itz80}
C. Itzykson and J.-B. Zuber, J. Math. Phys. {\bf 21},  411  (1980).

\bibitem{guh91}
T. Guhr, J. Math. Phys. {\bf 32},  336  (1991).

\bibitem{fyo03b}
Y. Fyodorov and E. Strahov, J. Phys. A {\bf 36},  3203  (2003).

\bibitem{Fyo03a}
Y. Fyodorov and J. Keating, J. Phys. A {\bf 36},  4035  (2003).

\bibitem{bre01}
E. Br\'ezin and S. Hikami, Commun. Math. Phys. {\bf 223},  363  (2001).

\bibitem{guh02b}
T. Guhr and H. Kohler, J. Math. Phys. {\bf 43},  2707  (2002).

\bibitem{mui82}
R. Muirhead, {\em Aspects of multivariate statistical theory}, {\em Probability
  and mathematical statistics} (John Wiley \& Sons, Inc., New York, 1982).

\bibitem{oko97}
A. Okounkov and G. Olshanski, Math. Res. Lett. {\bf 4},  69  (1997).

\bibitem{bor03}
A. Borodin and P. Forrester, J. Phys. A {\bf 36},  2963  (2003).

\bibitem{sch01d}
H. Schanze, E. Alves, C. Lewenkopf, and H.-J. St\"ockmann, Phys. Rev. E {\bf
  64},  065201(R)  (2001).

\bibitem{men03}
R. M\'ende{z-S\'a}nchez, U. Kuhl, M.~B.~C. Lewenkopf, and H.-J. St\"ockmann,
  Distribution of reflection eigenvalues in absorbing chaotic microwave
  cavities, cond-mat/0305090.

\bibitem{uji96a}
H. Ujino and M. Wadati, J. Phys. Soc. Jpn. {\bf 65},  653  (1996).

\bibitem{uji96b}
H. Ujino and M. Wadati, J. Phys. Soc. Jpn. {\bf 65},  2423  (1996).

\bibitem{uji97}
H. Ujino and M. Wadati, J. Phys. Soc. Jpn. {\bf 66},  345  (1997).

\end{thebibliography}
\end{document}